\documentclass[onecolumn]{article}
\usepackage{url}
\usepackage{amsmath}
\usepackage{bm}
\usepackage{dsfont}
\usepackage{amssymb}
\usepackage{multirow}
\usepackage[a4paper]{geometry}
\usepackage[utf8]{inputenc}
\usepackage{natbib}
\usepackage{pbox}
\usepackage{enumitem}
\usepackage{graphicx}
\usepackage{esvect}
\usepackage[export]{adjustbox}
\usepackage[figuresright]{rotating}
\usepackage{natbib}
\usepackage{pdflscape}
\usepackage{hyperref}

\usepackage{authblk}

\title{Identifying Brexit voting patterns in the British House of Commons: an analysis based on Bayesian mixture models with flexible concomitant covariate effects\thanks{This article has been accepted for publication in Journal of the Royal  Statistical  Society – Series C published by Oxford University Press. DOI: 10.1093/jrsssc/qlae004}}

\author{Marco Berrettini$^1$, Giuliano Galimberti$^1$\thanks{corresponding auhtor: giuliano.galimberti@unibo.it}, Saverio Ranciati$^1$, Thomas Brendan Murphy$^2$}
\affil{$^1$Department of Statistical Sciences, University of Bologna\\ $^2$School of Mathematics and Statistics, University College Dublin}
\date{\vspace{-5ex}}

\begin{document}




\maketitle




\begin{abstract}{Brexit and its implications are an ongoing topic of interest since the Brexit referendum in 2016. 
In 2019 the House of commons held a number of ``indicative'' and ``meaningful'' votes as part of the Brexit approval process. 
The voting behaviour of members of the parliament in these votes is investigated to gain insight into the Brexit approval process. 
In particular, a mixture model with concomitant covariates is developed to identify groups of members of parliament who share similar voting behaviour while also considering characteristics of the members of parliament. 
The novelty of the method lies in the flexible structure used to model the effect of concomitant covariates on the component weights of the mixture, with the (potentially nonlinear) terms represented as a smooth function of the covariates. 
Results show this approach allows to quantify the effect of the age of members of parliament, as well as preferences and competitiveness in the constituencies they represent, on their position towards Brexit. This helps grouping the aforementioned politicians into homogeous clusters, whose composition departs sensibly from that of the parties.
}
\end{abstract}

\noindent%
{\it Keywords:} P-splines, data augmentation, additive models, MCMC


%
%
\section{Introduction}
\label{sec:intro}

Statistical methodology can be a useful tool for the quantitative analysis of legislative voting records and, hence, has become widely present in the political science literature. In particular, clustering methods allow to identify heterogeneity in voting patterns even within political parties, as shown in \citet{spirling2010identifying} for the United Kingdom (UK) House of Commons. Although known for its discipline and cohesion, the British party system has always been wedged by the issue of European integration. This phenomenon has been thoroughly studied and is receiving even more interest, after the 2016 Brexit referendum. Nevertheless, to the best of the authors' knowledge, only a few papers have dealt with the MPs' voting behaviour on Brexit divisions from a quantitative point of view; see, e.g., \citet{aidt2021meaningful} and \citet{intal2021dissent}.

The empirical analysis described in this paper is based on data about the Parliamentary votes on Brexit, sometimes referred to as ``meaningful votes",  that are the parliamentary votes under the terms of Section 13 of the United Kingdom's European Union (Withdrawal) Act 2018. This Act requires the government of the United Kingdom (UK) to bring forward an amendable parliamentary motion at the end of the Article 50 negotiations between the government and the European Union (EU) to ratify the Brexit withdrawal agreement.

In the United Kingdom, a member of Parliament (MP) is an individual elected to serve in the House of Commons, the lower house of the Parliament of the United Kingdom. 
The Commons is an elected body consisting of 650 members representing constituencies (electoral areas). 
Among the MPs not taken into account in this study there are 4 Speakers (one Speaker and 3 deputies) 
who neither took part in the divisions nor voted, 7 members of the political party Sinn F{\'e}in which followed a policy of abstentionism (refusing to attend Parliament or vote on bills) and one MP who passed away on February 17th, 2019 and was replaced after April 4th, 2019.

At the end of March 2019, the government had not won any of the meaningful votes. 
This led to a series of non-binding ``indicative votes" on potential options for Brexit, and also to a delay of the departure date. 
The amendment tabled by Conservative MP Sir Oliver Letwin on March 25th, to take power to control business in the Commons away from the government on March 27th, to allow MPs to put forward business motions relating to Brexit, passed 329 -- 302.
Instead, the one tabled by Labour MP Dame Margaret Beckett was defeated 311 -- 314. 
This amendment would have required Parliament to vote favourably for a ``no deal" Brexit, or request an extension to Article 50 if the government was without a deal within seven days of leaving the European Union. 
The amended main motion (Letwin but not Beckett) passed 327 -- 300.
As a result of the Letwin amendment's success, indicative votes on Parliament's preferred Brexit options were held on March 27th. 
Eight propositions were voted upon, of which all eight failed:
\begin{itemize}
\item (B) No deal -- Conservative MP Mr John Baron's proposal to immediately leave the EU without any deal (For: 160, Against: 400);
\item (D) Common market 2.0 -- Conservative MP Mr Nicholas Boles' proposal to join the Single Market and a customs union (For: 189 -- against: 283);
\item (E) EFTA and EEA -- Conservative MP Mr George Eustice's proposal to remain in the Single Market outside of a customs union (For: 64 -- Against: 377);
\item (J) Customs union -- Conservative MP Mr Kenneth Clarke's proposal for a permanent customs union (For: 265 -- Against: 271);
\item (K) Labour's plan -- Labour's alternative position proposed by MP Mr Jeremy Corbyn, including a comprehensive customs union with the EU, close alignment with the Single Market, dynamic alignment on rights and protections, commitments on participation in EU agencies and funding programmes, and clear agreements on the detail of future security arrangements (For: 237 -- Against: 307);
\item (L) Revocation to avoid no deal -- Scottish National Party MP Ms Joanna Cherry's proposal's to revoke Article 50 (For: 184 -- Against: 293);
\item (M) Confirmatory public vote --  Labour MP Dame Margaret Beckett's proposal for a public vote on any withdrawal bill (For: 268 -- Against: 295);
\item (O) Managed no deal -- Conservative MP Mr Marcus Fysh's proposal to immediately leave the EU seeking a tariff-free trade agreement (For: 139 -- Against: 422).
\end{itemize}
\noindent It is worth noting that only the first and the last propositions express clear pro-leave positions, while the other 6 aim at mitigating the effects of Brexit, or even stopping it.

As Parliament had agreed to an extension of Article 50 to June 30th, the possibility of a third meaningful vote was raised and took place on March 29th, 2019. 
Mrs Theresa May promised she would resign as Prime Minister if the Withdrawal Agreement passed. 
In the end, Mrs Theresa May's deal was voted down again (For: 286 -- Against: 344), albeit by a smaller margin than in the previous two votes that took place on January, 15th, and March, 12th, respectively.
Further indicative votes were held on April 1st on four propositions chosen by the Speaker, all of which failed:
\begin{itemize}
\item (C) Customs union by Conservative MP Mr Kenneth Clarke (For: 273 -- Against: 276);
\item (D) Common market 2.0 by Conservative MP Mr Nicholas Boles (For: 261 -- Against: 282);
\item (E) Confirmatory public vote by Labour MPs Mr Peter Kyle and Mr Phil Wilson (For: 280 -- Against: 292);
\item (G) Parliamentary supremacy by Scottish National Party MP Ms Joanna Cherry (For: 191 -- Against: 292).
\end{itemize}
Note that all the four proposals were modified versions of those put to the vote on March 27th, although only the proposers of the confirmatory public vote and the title of Ms Cherry's one changed (previously, ``revocation to avoid no deal"). 
The proposal of a third round of indicative votes, to be held on April, 8th, was then rejected.
Due to huge opposition to the fourth withdrawal agreement, on May, 24th, Mrs Theresa May announced she would resign as Conservative Party leader and Prime Minister on June, 7th.

The remainder of the paper is organized as follows. The data used in the empirical analysis are introduced in Section~\ref{sec:2}; specification of the model is provided in Section~\ref{sec:3}, where also the associated Bayesian inference procedure is elicited; main results from the real data application are presented in Section~\ref{sec:6}, while the remaining ones, together with some simulation studies, are reported in the supplementary materials. Finally, Section~\ref{sec:7} is devoted to discussion and conclusions.

\section{Data}
\label{sec:2}
The main purpose of the analysis described in this paper is to identify groups of MPs whose opinions about Brexit, in terms of votes for the aforementioned divisions, are similar. 
In particular, $638$ members of the UK Parliament (MPs) are considered, whose voting data were collated using the ``{\fontfamily{pcr}\selectfont hansard}" R package \citep{hansard}. 
Furthermore, the influence of some concomitant information related to the MPs themselves, or the constituencies they represent, on group membership is considered. More specifically, $J=3$ concomitant covariates are included in the analysis: (i) age of the MP; (ii) share of Leave votes at the Brexit referendum in parliamentary constituencies; (iii) ``safeness" of the seat of each MP. 
Regarding the second covariate, it is worth noting the Brexit referendum vote was not counted by constituency, except in Northern Ireland. Some local councils (districts) republished local results by electoral ward or constituency. Some constituencies are coterminous with (overlap) their local government district. For the others, \citet{hanretty2017areal} estimated through a demographic model the Leave and Remain vote shares. 

About the third covariate, \citet{apostolova2017general} analyzed and made available the results of the 2017 UK general election and, in particular, the number of votes taken by each party for each of the 650 constituencies. 
In this dataset, 12 main parties are considered, while all the others are gathered together, unless one of these won the seat: the votes taken by the winning party are counted separately and placed in another category, for a total of $P=13$ categories. 
To quantify how much an MP, or the party they represent, was supported in the constituency they were elected into, a measure of the degree of heterogeneity of votes among parties in that constituency is considered. In particular, by denoting with $\omega_{pi}$ the share of votes taken by party $k\, (k=1,\dots,P)$, the entropy \citep{shannon1948entropy} of the votes in the $i$-th constituency can be computed as:
\begin{equation*}
\mbox{EN}(\boldsymbol{\omega}_{i})=-\sum_{k=1}^P \omega_{ki}\log(\omega_{ki}).
\end{equation*}
In this case, the entropy $\mbox{EN}(\boldsymbol{\omega}_{i})$ quantifies the uncertainty in predicting the number of votes taken by a party drawn at random in a given constituency. It ranges from a minimum of 0, which corresponds to a situation of no heterogeneity (i.e. all votes are taken by a single party), to a maximum of $\log (P)$ when there is equidistribution of votes between the parties.
Thus, by considering $\exp[\mbox{EN}(\boldsymbol{\omega}_{i})]$, a quantity that can be interpreted as the effective number of competing political parties (or candidates) in a given constituency is obtained, in analogy with the ``effective number of species" introduced in the biological literature by \citep{macarthur1965patterns}.

\section{Methodology}
\label{sec:3}
\subsection{Methods from the literature}
Mixture models are the basis of many model-based clustering methods, where uncertainty can be accounted for in a probabilistic framework. Fields in which mixture models have been successfully applied include agriculture, astronomy, bioinformatics, biology, economics, engineering, genetics, imaging, marketing, medicine, neuroscience, physics psychiatry, psychology and social sciences. Extensive reviews of mixture models and their application are given in \citet{everitt1981finite}, \citet{titterington1985statistical}, \citet{mclachlan1988mixture}, \citet{lindsay1995mixture}, \citet{bohning1999computer}, \citet{mclachlan2004finite}, \citet{fruhwirth2006finite}, \citet{mengersen2011mixtures}, \citet{mcnicholas2016mixture} and \citet{bouveyron2019model}.

Mixtures of experts (MoE) models provide a way to extend mixture models, allowing the parameters to depend on concomitant covariate information. This nomenclature arises in \citet{jacobs1991adaptive}, whose authors model the prior probabilities of latent group membership (i.e. the component weights) as a logistic function of the covariates. However, some models belonging to the general class of MoE were already present in the statistical literature under the name of switching regression models \citep{quandt1972new}, concomitant variable latent-class models \citep{dayton1988concomitant}, clusterwise regression models \citep{desarbo1988maximum} and mixed models \citep{wang1996mixed}. 
Any mixture model which incorporates covariates or concomitant variables falls within the family of MoE models, which ranges from the ``full" MoE model, in which all model parameters are functions of covariates, to the special cases where some of the model parameters do not depend on covariates. A graphical model representation of the suite of MoE models is provided by \citet{murphy2019gaussian}.
The MoE framework 
allows for a wide range of applications: among others, rank data \citep{gormley2008mixture}, network data \citep{gormley2010mixture}, time series data \citep{fruhwirth2012labor}, non-normal data \citep{chamroukhi2015non, nguyen2016laplace} and longitudinal data \citep{tang2016mixture} have been developed. Recent reviews of MoE models is provided by \citet{nguyen2018practical} and \cite{gormley2019mixture}.

\citet{jacobs1991adaptive} and \citet{jordan1994hierarchical} derive maximum likelihood (ML) estimates for MoE models via the expectation-maximization (EM) algorithm. However, the EM algorithm for fitting MoE models is often difficult in practice, usually due to a complex component density and/or component weights model or a large parameter set. 
Inference for MoE models can also be carried out within the Bayesian paradigm, using a Markov chain Monte Carlo (MCMC) algorithm. Bayesian approaches for mixture models are summarized by \citet{fruhwirth2006finite}. Both the Gibbs sampler \citep{geman1984stochastic} and the Metropolis-Hastings (MH) \citep{metropolis1953equation} algorithm are typically required, though sampling the component weight parameters through a MH-algorithm brings issues, such as choosing suitable proposal distributions and tuning parameters. 
Alternatively, \citet{fruhwirth2012labor} introduced data augmentation of the multinomial logit regression model, based on the difference random utility model (dRUM) representation, in the context of MoE models. Following \citet{fruhwirth2010data}, the authors approximate the logistic distribution by a finite scale mixture of Gaussian distributions in order to avoid any MH step. 
%

This paper focuses on the gating network MoE model \citep{murphy2019gaussian}, where only the component weights depend on concomitant covariates. In particular, a more flexible specification of these covariate effects is considered, in terms of a sum of smooth functions as in a generalized additive model \citep{hastie1990generalized}. Several proposals are available for modeling and estimating the smooth functions, see, e.g., \citet{hastie2009elements} and \citet{wood2017generalized} for an overview. 
In order to achieve a parsimonious representation of these smooth functions Bayesian P-splines are used, as suggested by \citet{lang2004bayesian}. Following \citet{fruhwirth2012labor}, data augmentation is exploited based on the dRUM scheme, thus introducing a set of auxiliary variables.

\subsection{Model specification}
Consider an independent and identically distributed sample of 
outcome observations $\{\mathbf{y}_i\}$, with $i = 1, \dots , n$, from a population clustered by a $G$ components finite mixture model. Each component $g$, for $g = 1,\dots,G$, is modelled by the probability (density) function $f(\mathbf{y}_i|\boldsymbol{\theta}_g)$  with parameters $\boldsymbol{\theta}_g$ and weight $p_g$ , such that $0<p_g<1$ and $\sum_{g=1}^G p_g=1$. All component densities $f(\mathbf{y}|\boldsymbol{\theta}_g)$, for $g=1,\dots, G$, are assumed to arise from the same parametric distribution family, depending on the specific nature of the outcome variable.
For the discrete case, one writes $y_{iqc}=1, c=1,\dots,C_q,$ if the $i$-th unit presents the $c$-th category out of $C_q$ possibilities for the $q$-th variable, $q=1,\dots,Q$. For each group $g$, conditional independence is assumed between these $Q$ manifest variables, as in \citet{spirling2010identifying} leading to the following multinomial distribution:
\begin{equation*}
f(\mathbf{y}_i|\boldsymbol{\theta}_g)=\prod_{q=1}^Q\prod_{c=1}^{C_q}(\theta_{gqc})^{y_{iqc}}.
\end{equation*}
\noindent In the Brexit data application, vector $\boldsymbol{\theta}_{gq}=(\theta_{gq1},\theta_{gq2},\theta_{gq3})$ contains the probabilities of all the possible outcomes (``absent", ``aye" and ``no" respectively) for each of the $Q=16$ divisions and each group $g=1,\dots,G$.

Observation $i$ has $J$ associated covariates 
$\mathbf{x}_i=$ $(1,$ $x_{i1},$ $\dots,$ $x_{iJ^*-1},$ $x_{iJ^*},$ $x_{iJ^*+1},$ $\dots,$ $x_{iJ})$, of which the last $J-J^*$ are metrical, with $J^*\in [1,J]$.
The gating network MoE model extends the finite mixture model by allowing the distribution of the latent variable to depend on the concomitant variables:
\begin{equation}
f(\mathbf{y}_i|\mathbf{x}_i)=\sum_{g=1}^G p_g(\mathbf{x}_i)f(\mathbf{y}_i|\boldsymbol{\theta}_g).
\label{eq:01}
\end{equation}

\noindent \citet{jacobs1991adaptive} model the components' weights using a multinomial logit regression model. Arbitrarily selecting a ``reference" class - for example, the $G$-th one, one can assume that the log-odds of the latent class (prior) membership $p_g$, with respect to that class $G$, are linear functions of the covariates:
\begin{equation*}
\log\, \frac{p_g(\mathbf{x}_i)}{p_G(\mathbf{x}_i)}=\mathbf{x}_i\boldsymbol{\gamma}_g,\quad g=1,\dots,G-1,
\end{equation*}
\noindent where $\boldsymbol{\gamma}_g, \,g=1,\dots,G-1,$ denotes the vector of coefficients corresponding to the $g$-th latent class.
Following some simple algebra, this produces the general result that:
\begin{equation*}
p_{g}(\mathbf{x}_i)=\frac{\exp(\mathbf{x}_i\boldsymbol{\gamma}_g)}
{1+\sum_{g=1}^{G-1}\exp(\mathbf{x}_i\boldsymbol{\gamma}_g)}.
\end{equation*}

\noindent The main goal of this paper is to introduce extra-flexibility in the effects of the covariates on $p_g(\mathbf{x})$ by extending the linear predictor according to the additive paradigm (see, for example, \citet{green1993nonparametric} or \citet{hastie2009elements}). More precisely, the linear predictor is defined as:
\begin{equation*}
\ln \frac{p_g(\mathbf{x}_i)}{p_G(\mathbf{x}_i)}=\eta_{gi}
=\sum_{j=0}^{J^*}\gamma_{gj}x_{ij}+\sum_{j=J^*+1}^{J}s_{gj}(x_{ij}).
\end{equation*}
\noindent Here $s_{g,J^*+1},\dots,s_{gJ}$ are unknown smooth functions of the metrical covariates.
These smooth functions can be expressed as:
\[s_{gj}(x_{ij})=\sum_{\rho=1}^{m}\beta_{gj\rho}B_{j\rho}(x_{ij}),\]
were $\boldsymbol{\beta}_g$ denotes the vector containing the parameters associated to the nonlinear part of the predictor for the $g$-th component ($g=1,\dots,G-1$), and $B_{j\rho}(\cdot)$ is a B-spline basis function for a cubic spline ($j=J^*+1,\dots,J$ and $\rho=1,\dots,m$) . 
To ensure identifiability of the additive predictors, each function $s_{gj}(\mathbf{x}_j)$ is constrained to have zero means, that is, 
\begin{equation*}
\frac{1}{\mbox{range}(\;{x}_j)}\int s_{gj}(x_j)\, \mbox{d}x_j=0, \quad j=J^*+1, \dots, J.
\label{constr}
\end{equation*}
\noindent This can be incorporated into estimation via iterative algorithms by centering the functions $s_{gj}(\mathbf{x}_j)$ about their means in every iteration and thus adding the means to the intercept $\mathbf{\gamma}_{g0}$. See Section \ref{sec:4.1} for further details about Bayesian constrained sampling.

By defining the $n \times m$ design matrices $\mathbf{B}_j$, where the element in row $i$ and column $\rho$ is given by $B_{j\rho}(x_{ij})$, the predictor can be rewritten in matrix notation:
\begin{equation*}
\boldsymbol{\eta}_g=\mathbf{X}\boldsymbol{\gamma}_g+\sum_{j=J^*+1}^{J}\mathbf{B}_{j}\boldsymbol{\beta}_{gj},
\end{equation*}
where $\boldsymbol{\beta}_{gj}=(\beta_{gj1},\dots, \beta_{gj1})', j=J^*+1,\dots, J$, correspond to the vectors of unknown regression coeffcients. The matrix $\mathbf{X}$ is the design matrix of fixed effects. 

\subsection{Bayesian Inference}
\label{sec:4}

\citet{lang2004bayesian} propose to set a high number of knots to ensure enough flexibility, and define the priors for the regression parameters $\beta_{gj}$ in terms of a random walk:
\begin{equation}
\beta_{gj\rho}=\beta_{gj,\rho-1}+u_{gj\rho}, \quad u_{gj\rho}\sim \mbox{N}(0,\tau_{gj}^2).
\label{eq:03}
\end{equation}
\noindent The amount of smoothness is controlled by the additional variance parameters $\tau_{gj}^2$, which correspond to the inverse smoothing parameters in the frequentist approach; this parameter $\tau_{gj}^2$ protects against possibile overfitting if a large number of knots is chosen.
The priors can be equivalently written in the form of global smoothness priors
\begin{equation*}
\boldsymbol{\beta}_{gj}|\tau_{gj}^2\propto \exp \left(-\frac{1}{2\tau_{gj}^2}\boldsymbol{\beta}_{gj}'\mathbf{K}_{j}\boldsymbol{\beta}_{gj}\right),
\end{equation*}
\noindent where the penalty matrix $\mathbf{K}_j$ is given by $\mathbf{K}_j=\boldsymbol{\Delta}_1'\boldsymbol{\Delta}_1$, with $\boldsymbol{\Delta}_1$ being the first order difference matrix \citep[][Chapter 2, p. 52]{rue2005gaussian}. Because $\mathbf{K}_{j}$ is rank deficient with $\mbox{rank}(\mathbf{K}_{j}) = m-1$ for a first-order random walk, the prior is improper. It is worth mentioning that these kind of models are usually referred to in the literature as Intrinsic Gaussian Markov Random Fields \citep{rue2005gaussian}. \citet{lang2004bayesian} suggest a number of knots between 20 and 40.
Hyperpriors are assigned to the variances $\tau_{gj}^2$, selecting Inverse Gamma distributions $\tau_{gj}^2\sim \mbox{IG} (a_{gj},b_{gj})$, with $a_{gj} = 1$ and a small value for $b_{gj}$, for example, $b_{gj}=0.005$ leading to almost diffuse priors for  $\tau_{gj}^2$. Regarding the coefficients involved in $\eta_{gi}$, Gaussian priors are assumed for the fixed effects parameters: $\boldsymbol{\gamma}_g\sim\mbox{N}(\mathbf{0}_{J^*+1},v\,\mathbf{I}_{J^*+1})$, with variance hyperparameter $v$ set high (e.g. equal to 100), enough to make the prior non informative.

Following \citet{fruhwirth2012labor}, any multinomial logit model can be written as a binary one in the partial dRUM representation:
\begin{equation}
z_{gi}=\eta_{gi}-\log\left(\sum_{l\neq g}\lambda_{li}\right)+\epsilon_{gi}, \quad
D_{gi}=\mathds{1} (z_{gi}>0) ,
\label{eq:04}
\end{equation}
\noindent where $\mathds{1}$ denotes the indicator function, $z_{gi}$ is a latent variable, $\lambda_{gi}=\exp(\eta_{gi})$ and $\epsilon_{gi}$ are i.i.d. errors following a logistic distribution. 

Given $\lambda_{1i},\dots,\lambda_{Gi}$ and the latent indicator variables $D_{1i},\dots,D_{Gi}$, the latent variables $z_{1i},\dots, z_{Gi}$ are distributed according to an Exponential distribution and can be easily sampled in a data augmented implementation.
To avoid any Metropolis-Hastings step, \citet{fruhwirth2010data} approximate, for each $\epsilon_{gi}$, the logistic distribution by a finite scale mixture of H normal distributions with zero means and variances $\omega_{1},\dots,$ drawn with fixed probabilities $w_1,\dots,w_H$. The same authors obtained their finite scale mixture approximation by minimizing the Kullback-Leibler divergence between the densities, and recommend choosing $H = 3$ in larger applications, where computing time matters, and to work with $H = 6$ whenever possible. In a second step of data augmentation, the component indicator $r_{gi}$, taking value $h=1,\dots, H$, is introduced as yet another latent variable. Conditional on the latent variables $\mathbf{z}_g$ and the indicators $\mathbf{r}_g = (r_{g1}, \dots , r_{gn})$, the binary logit regression model \eqref{eq:04}reduces to a Gaussian regression model.

Regarding the parameters of each component, appropriate full conditionals can be exploited in order to obtain samples from the posterior distribution. For the data application, a Dirichlet distribution is assumed for the probabilities of each division's possible outcome $\boldsymbol{\theta}_{gq}$ (for $g=1,\dots, G;\, q=1,\dots, Q$), with hyperparameters $\delta_1,\dots,\delta_{C_q}$ all set equal to 1.
\subsection{MCMC Algorithm}
\label{sec:4.1}
Based on the representation of Section \ref{sec:4}, a new MCMC algorithm is implemented for fixed $G$ by integrating the scheme proposed by \citet{fruhwirth2012labor}, with the Bayesian P-spline approach by \citet{lang2004bayesian}. A sketch of the algorithm is comprised of the following steps:
\begin{enumerate}

\item sample the regression coefficients $\boldsymbol{\beta}_g$ conditional on $\mathbf{z}_{g}$ and $\mathbf{r}_g$, for $g=1,\dots, G-1$. Using the prior in \eqref{eq:03}, the conditional posterior of $\boldsymbol{\beta}_{gj}$ is given by a multivariate normal density. Straightforward calculations \citep{brezger2006generalized} show that the precision matrix $\mathbf{P}_{gj}$ and the mean $\mathbf{m}_{gj}$ of $\boldsymbol{\beta}_{gj}|\cdot$ are given by
\begin{equation*}
\begin{split}
&\mathbf{P}_{gj}=\mathbf{B}_j'\mathbf{W}_g^{-1}\mathbf{B}_j+\frac{1}{\tau_{gj}^2}\mathbf{K}_{j}, \\
&\mathbf{m}_{gj}=\mathbf{P}_{gj}^{-1}\mathbf{B}_j'\mathbf{W}_g^{-1}\left(\mathbf{z}_g-\boldsymbol{\tilde{\eta}}_{g,-j}+\log\sum_{l\neq g}\boldsymbol{\lambda}_{l}\right),
\end{split}
\end{equation*}
\noindent where $\tilde{\boldsymbol{\eta}}_{g,-j}$ is the part of the predictor associated with all but the $j$-th effects in the model, and $\mathbf{W}_g$ is a $n\times n$ diagonal matrix with nonzero elements equal to the randomly drawn variances $\omega_{1g}=s_{r_{g1}}^2,\dots, \omega_{ng}=s_{r_{gn}}^2$ for the $g$-th group;

\item center each smooth function $s_{gj}(\mathbf{x}_j)$: imposing the constraint in \eqref{constr} is equivalent to sampling $\boldsymbol{\beta}_{gj}|(\mathbf{1}_n'\mathbf{B}_j\boldsymbol{\beta}_{gj}=\mathbf{0})$ for each $j=J^*+1,\dots,J$. Following Algorithm 2.6 in \citet{rue2005gaussian}, this can be done by trasforming each vector of coefficients $\boldsymbol{\beta}_{gj}$ as follows:
\begin{equation*}\label{eq:ruwheldcentr}
\boldsymbol{\tilde{\beta}}_{gj}=\boldsymbol{\beta}_{gj}-\mathbf{P}_{gj}^{-1}\mathbf{B}_j'\mathbf{1}_n\left(
\mathbf{1}_n'\mathbf{B}_j \mathbf{P}_{gj}^{-1}\mathbf{B}_j'\mathbf{1}_n
\right)^{-1}\mathbf{1}_n'\mathbf{B}_j\boldsymbol{\beta}_{gj};
\end{equation*} 

\item sample the fixed effects parameters $\boldsymbol{\gamma}_g$ from a multivariate normal density with precision matrix $\mathbf{P}_{\boldsymbol{\gamma}_g}$ and the mean vector $\mathbf{m}_{\boldsymbol{\gamma}_g}$:
\begin{equation*}
\begin{split}
&\mathbf{P}_{\boldsymbol{\gamma}_g}=\mathbf{X}'\mathbf{W}_g^{-1}\mathbf{X}+v^{-1}\mathbf{I}_{J^*+1},\\
&\mathbf{m}_{\boldsymbol{\gamma}_g}=\mathbf{P}_{\boldsymbol{\gamma}_g}^{-1}\mathbf{X}'\mathbf{W}_g^{-1}\left(\mathbf{z}_g-\boldsymbol{\tilde{\eta}_{g,-\gamma}}+\log\sum_{l\neq g}\boldsymbol{\lambda}_{l}\right),
\end{split}
\end{equation*}
\noindent where $\boldsymbol{\tilde{\eta}_{g,-\gamma}}$ represents the nonlinear part of the $g$-th predictor;

\item sample the variance parameters $\tau_{gj}^2$ conditional on $\boldsymbol{\tilde{\beta}}_{gj}$:
\begin{equation*}
\tau^2_{gj}|\boldsymbol{\beta}_{gj} \sim \mbox{IG}\left(a_{gj}+ \frac{\mbox{rank}(\mathbf{K}_{j})}{2}, b_{gj}+ \frac{1}{2}\boldsymbol{\tilde{\beta}}_{gj}'\mathbf{K}_{j}\boldsymbol{\tilde{\beta}}_{gj}\right);
\end{equation*}

\item sample all utilities $z_{1i},\dots, z_{G-1,i}$ simultaneously for each $i$ from:
\begin{equation*}
z_{gi}=\log\left(\frac{\lambda_{gi}}{\log\sum_{l\neq g}\lambda_{li}}U_{gi}+D_{gi}\right)
-\log\left(1-U_{gi}+\frac{\lambda_{gi}}{\log\sum_{l\neq g}\lambda_{li}}D_{gi}\right),
\end{equation*}
\noindent with $U_{gi}\sim \mbox{Unif}(0,1);$

\item sample the component indicators $r_{gi}$ conditional on $z_{gi}$ from:
\begin{equation*}
\mbox{Pr}(r_{gi}=h|z_{gi},\boldsymbol{\beta}_g,\boldsymbol{\gamma}_g)\propto\frac{w_h}{s_h}\exp\left[-\frac{1}{2}\left(\frac{z_{gi}-\eta_{gi}+\log\sum_{l\neq g}\lambda_{li}}{s_h}\right)^2\right];
\end{equation*}

\item sample the cluster-specific conditional probabilities $\boldsymbol{\theta}_{g1},\dots,\boldsymbol{\theta}_{gQ}$ given the allocation vector $\mathbf{D}_g=(D_{g1},\dots,D_{gn})'$, from a total of $G\cdot Q$ Dirichlet distributions:
\begin{equation*}
\boldsymbol{\theta}_{gq}|\mathbf{D}_g,\mathbf{y}\sim \mbox{Dir}\left(\delta_1+\sum_{i=1}^n D_{gi}\cdot y_{iq1},\dots,\delta_{C_q}+\sum_{i=1}^n D_{gi}\cdot y_{iqC_q}\right);
\end{equation*}

\item classify each individual $i$ according to Bayes' rule: draw the allocation vector $\mathbf{D}_i=(D_{1i},\dots,D_{Gi})'$ from the following discrete probability distribution which combines the likelihood $f(\mathbf{y}_i|\boldsymbol{\theta}_g)$ and the prior $p_{g}(\mathbf{x}_i)$:
\begin{equation*}
\mbox{Pr}(D_{gi}=1|\mathbf{y}_i,\mathbf{x}_i,\boldsymbol{\beta},\boldsymbol{\gamma},\boldsymbol{\theta})
\propto\frac{\exp(\eta_{gi})}
{1+\sum_{g=1}^{G-1}\exp(\eta_{gi})}
\prod_{q=1}^{Q}\prod_{c=1}^{C_q}(\theta_{gqc})^{y_{iqc}}.
\end{equation*}

\end{enumerate}
\subsection{Posterior inference and model selection}
\label{sec:2.2.4}
Once the MCMC algorithm has completed the prefixed number $T$ of iterations, posterior inference is carried out by estimating each parameter's posterior mean over the last $T-T_0$ draws of the chains, with $T_0$ defining the burn-in phase. Posterior quantities can be computed for the smooth functions by considering them as linear combinations of spline bases and the corresponding regression coefficients' estimates. The uncertainty associated to the smooth functions is quantified via their pointwise percentiles (usually 2.5 - 97.5 or 5 - 95), for each function over the last $T-T_0$ posterior draws. 
 
Observations can be allocated into the $G$ components using the maximum-a-posteriori (MAP) rule. In particular, each unit $i=1,\dots,n$ is assigned to the component $\hat{c}_i$ such that
\begin{equation} 	
\hat{c}_i=\arg \max_g\left(\sum_{t=T_0}^T D_{1i}^{(t)},\dots, \sum_{t=T_0}^T D_{Gi}^{(t)}\right).
\end{equation} 
\noindent where $\mathbf{D}_i^{(t)}=(D_{1i}^{(t)},\dots,D_{Gi}^{(t)})$ represents the allocation vector for unit $i$ at iteration $t$.
Sometimes, using the MAP rule, one or more components could have no units assigned to them: thus, it might be worth distinguishing between the number of components $G$ and the number of non-empty components, denoted as
\begin{equation}\label{eq:nonempty}
\tilde{G} =\sum_{g=1}^G\mathds{1}\left(\sum_{i=1}^n \mathds{1}(\hat{c}_{i}=g)>0\right).
\end{equation}
\noindent Choosing the number of components in a mixture model is an important problem, which originated many efforts in the statistical literature. 
In most approaches, selecting $G$ is related to the number of free parameters, which is not clear for the proposed model, due to the presence of regulariziation induced by the prior distribution on the regression coefficients.
One simple solution in the Bayesian framework is given by the AICM \citep{RafEtAl2007}, whose formula depends only on the log-likelihoods from the posterior simulation:
\begin{equation}
\mbox{AICM}=2(\bar{l}-s_l^2),
\end{equation}
\noindent where $\bar{l}$ and $s_l^2$ are the sample mean and variance of the sequence of log-likelihoods $f(\mathbf{y}_i|\boldsymbol{\theta}_{\mathbf{D}_i}^{(t)})$, for each iteration $t=T_0,\dots,T$, after the burn-in. 
AICM has already been applied successfully in the mixture modelling context, for example by \citet{erosheva2007describing}, \citet{gormley2010mixture}, \citet{gormley2011mixture}, and \citet{mollica2017bayesian}.

\subsection{Label switching}

As for any finite mixture model, label switching may occur during MCMC sampling; see \citet[][Section 3.5]{fruhwirth2006finite} for a review. 
To identify a mixture of experts model, \citet{fruhwirth2012labor} suggest to focus on a subset of a group-specific parameter and apply $k$-means clustering (with $G$ clusters) to the posterior draws. 
MCMC draws belonging to the same group are assigned to the same cluster by $k$-means clustering, and the resulting classification sequences $\zeta_t=(S_1^{(t)},\dots,S_G^{(t)})$ -- where each $S_g^{(t)}$, $g=1,\dots,G$, is a classification index taking values in \{1,\dots,G\} -- show how to re-arrange the group-specific parameters for each iteration $t=1,\dots,T$, even if label switching occurred during sampling. In particular, if the mixture is not overfitting the number $G$ of groups, $\zeta_t$ is a permutation of $\{1,\dots,G\}$, and a unique labeling is achieved by reordering the draws in the following way:
\begin{itemize}
\item relabel the hidden allocations $\mathbf{D}_1,\dots,\mathbf{D}_n$ through the inverse $\zeta_t^{-1}$: re-arrange $\mathbf{D}_{1},\dots,\mathbf{D}_{G}$ by $\mathbf{D}_{\zeta_t^{-1}(1)},\dots,\mathbf{D}_{\zeta_t^{-1}(G)}$, respectively;
\item relabel the group-specific parameters through $\zeta_t^{-1}(1)$, $\dots$, $\zeta_t^{-1}(G)$: re-arrange $\boldsymbol{\theta}_1,\dots,\boldsymbol{\theta}_G$ by $\boldsymbol{\theta}_{\zeta_t^{-1}(1)}, \dots, \boldsymbol{\theta}_{\zeta_t^{-1}(G)}$;
\item relabel the regression coefficients $\boldsymbol{\gamma}$ and $\boldsymbol{\beta})$, corresponding to the linear and nonlinear part of the predictor in the multinomial logistic regression model: substitute $\boldsymbol{\gamma}_1,\dots,\boldsymbol{\gamma}_G$ by $\boldsymbol{\gamma}_{\zeta_t^{-1}(1)}-\boldsymbol{\gamma}_{\zeta_t^{-1}(G)}$, $\boldsymbol{\gamma}_{\zeta_t^{-1}(2)}-\boldsymbol{\gamma}_{\zeta_t^{-1}(G)}$, $\dots$, $\boldsymbol{\gamma}_{\zeta_t^{-1}(G)}-\boldsymbol{\gamma}_{\zeta_t^{-1}(G)}=\mathbf{0}$ and, thanks to the fact that the additive predictors are still linear in the parameters, substitute $\boldsymbol{\beta}_1,\dots,\boldsymbol{\beta}_G$ by $\boldsymbol{\beta}_{\zeta_t^{-1}(1)}-\boldsymbol{\beta}_{\zeta_t^{-1}(G)},\boldsymbol{\beta}_{\zeta_t^{-1}(2)}-\boldsymbol{\beta}_{\zeta_t^{-1}(G)},\dots,\boldsymbol{\beta}_{\zeta_t^{-1}(G)}-\boldsymbol{\beta}_{\zeta_t^{-1}(G)}=\mathbf{0}$. Subtracting $\boldsymbol{\gamma}_{\zeta_t^{-1}(G)}$ and $\boldsymbol{\beta}_{\zeta_t^{-1}(G)}$, respectively, from all draws ensures that the regression coefficients of the baseline are all equal to 0 in the identified model.
\end{itemize}

\section{Uncovering heterogeneity in MPs' voting behaviour}
\label{sec:6}

\begin{figure}
\includegraphics[width=\textwidth]{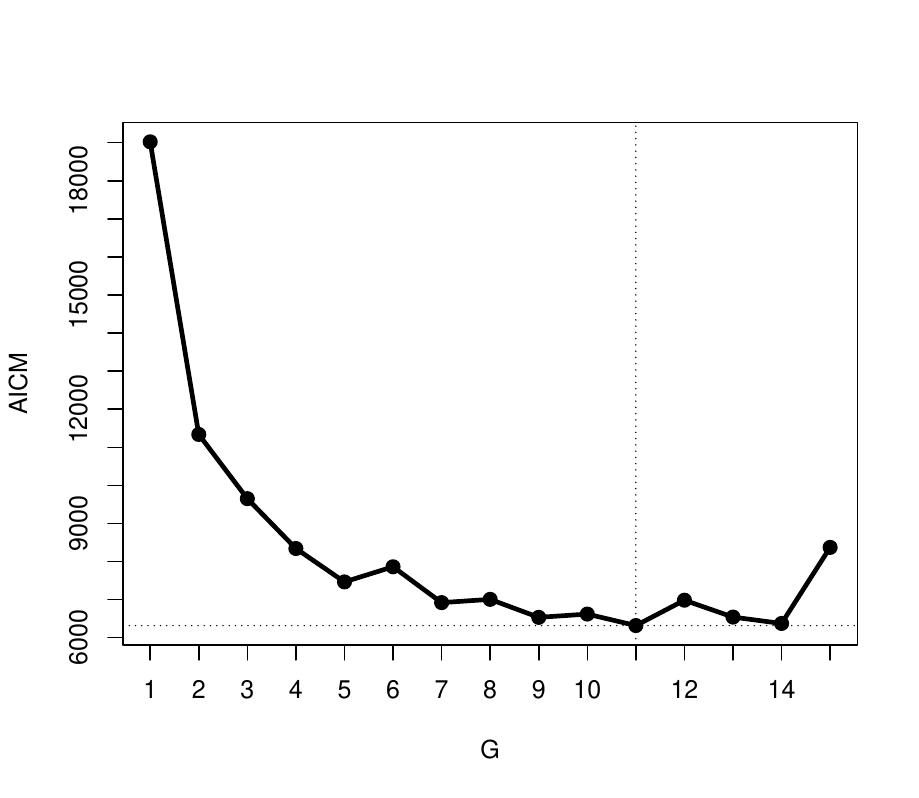}
\caption{AICM values corresponding to different number of components $G$ for the semiparametric gating network mixture of experts model} 
\label{fig:AICM}       

\end{figure}

To identify groups of MPs with similar opinion about Brexit, values of $G$ ranging from 1 to 15 are considered for the semiparametric MoE model described in Section \ref{sec:3}. 
For each value of $G$, 4000 MCMC draws are considered after a burn-in of 1000 draws. 
The optimal number of components suggested by the AICM is 11, although Figure \ref{fig:AICM} shows that the AICM curve is quite flat between 9 and 14.
Cluster composition in terms of political party membership is shown in Table~\ref{tab:clus}. 
For space reasons, only the results relative to the most meaningful clusters detected in the analysis are described. 

\begin{table}
\caption{Posterior allocation and political party membership of the $n=638$ MPs. Adjusted Rand Index: 0.458. 
(C = Conservatives, DUP = Democratic Unionist Party, GP = Green Party, Ind = Independents, Lab = Labour, LD = Liberal Democrats, PC = Plaid Cymru, SNP = Scottish National Party). \label{tab:clus}} 

\begin{tabular}{l|ccccccccccc}
\hline
   & \multicolumn{11}{c}{Cluster}\\
  Party   & 1 &  2 &  3 &  4 &  5 &  6 &  7 &  8 &  9 & 10 & 11 \\
  \hline
    C    & 0 & 21 & 16  & 5 & 21  & 0 & 33  & 3  & 3  & 0 & 211 \\
  DUP  & 0  & 0  & 0 & 10  & 0  & 0  & 0 &  0  & 0  & 0  & 0 \\
  GP   & 0  & 0  & 0  & 0  & 0  & 0  & 0  & 0  & 1  & 0  & 0 \\
  Ind  & 0  & 0  & 0  & 0  & 0  & 1  & 0  & 2 & 12  & 0 &  1 \\
  Lab & 19  & 1 &  4  & 0  & 0 &108  & 1 &  2 &  6 & 101 &  5 \\
  LD   & 0  & 1  & 0  & 0  & 0  & 1  & 0  & 9  & 1  & 0  & 0 \\
  PC   & 0  & 0  & 0  & 0  & 0  & 0  & 0  & 4  & 0  & 0  & 0 \\
  SNP  & 0 &  0  & 0 &  0 &  0 &  0 &  1  & 34 & 0 &  0 &  0 \\
   \hline
\end{tabular}

\end{table}
\begin{figure}

\includegraphics[width=\textwidth]{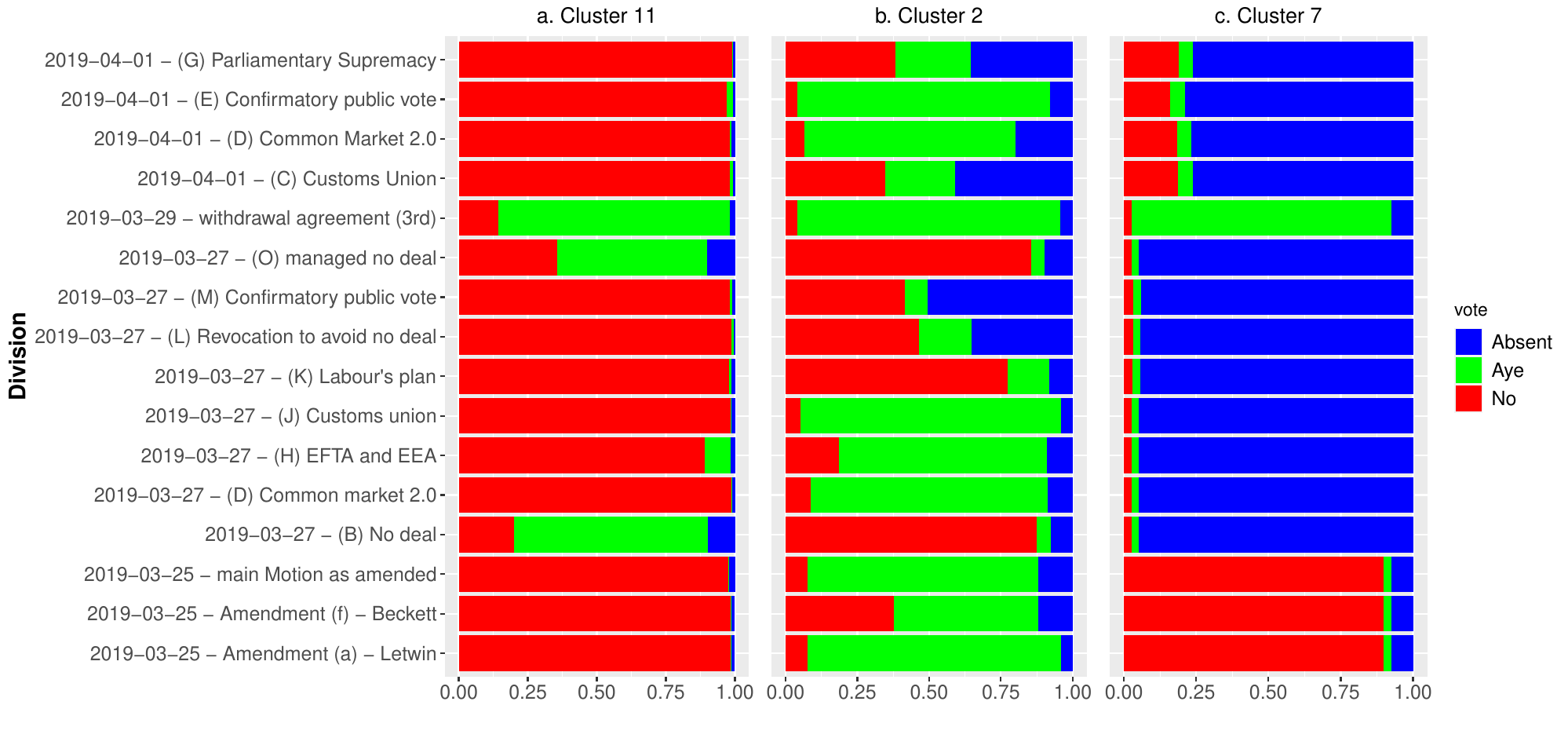}\\
\includegraphics[width=\textwidth]{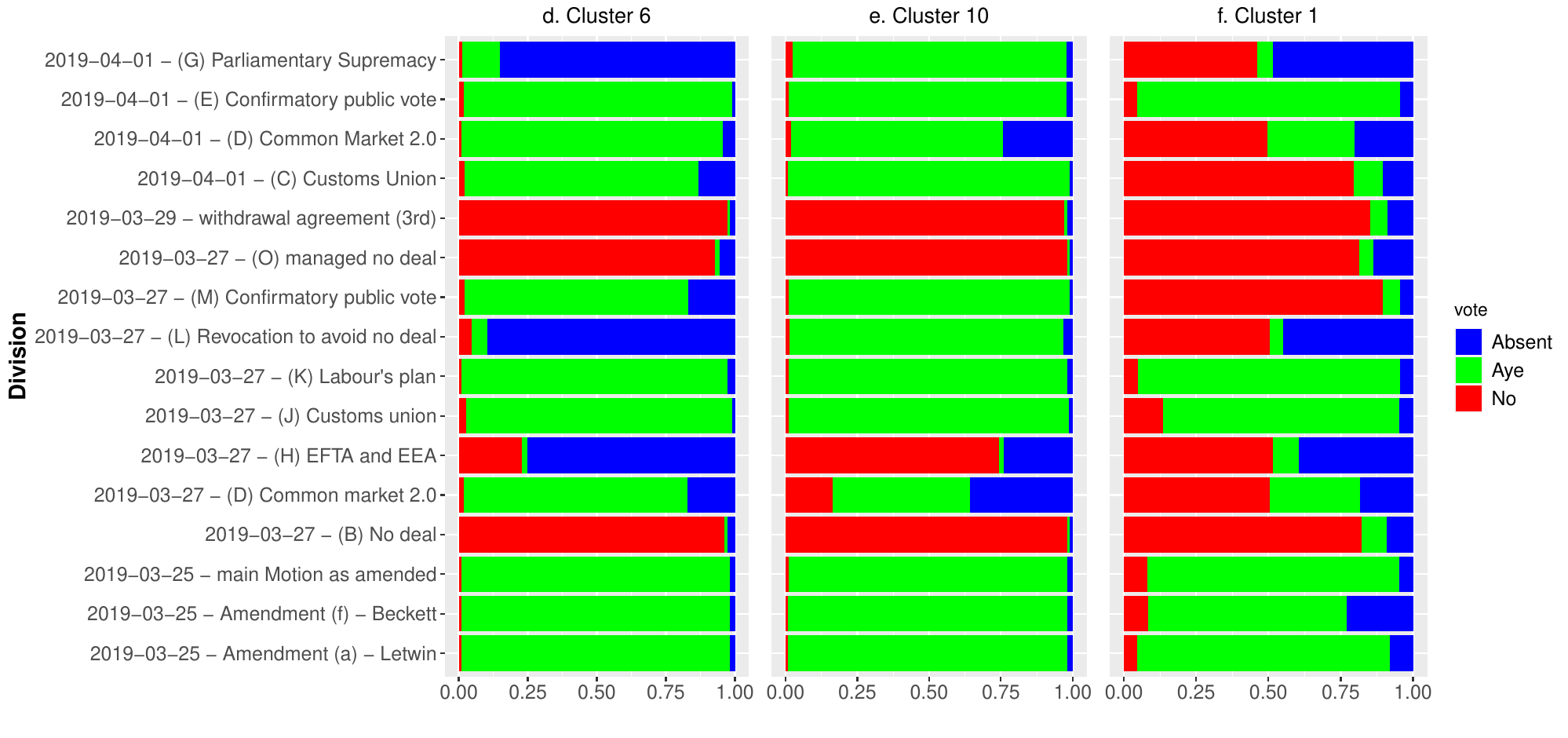}\\
\includegraphics[width=\textwidth]{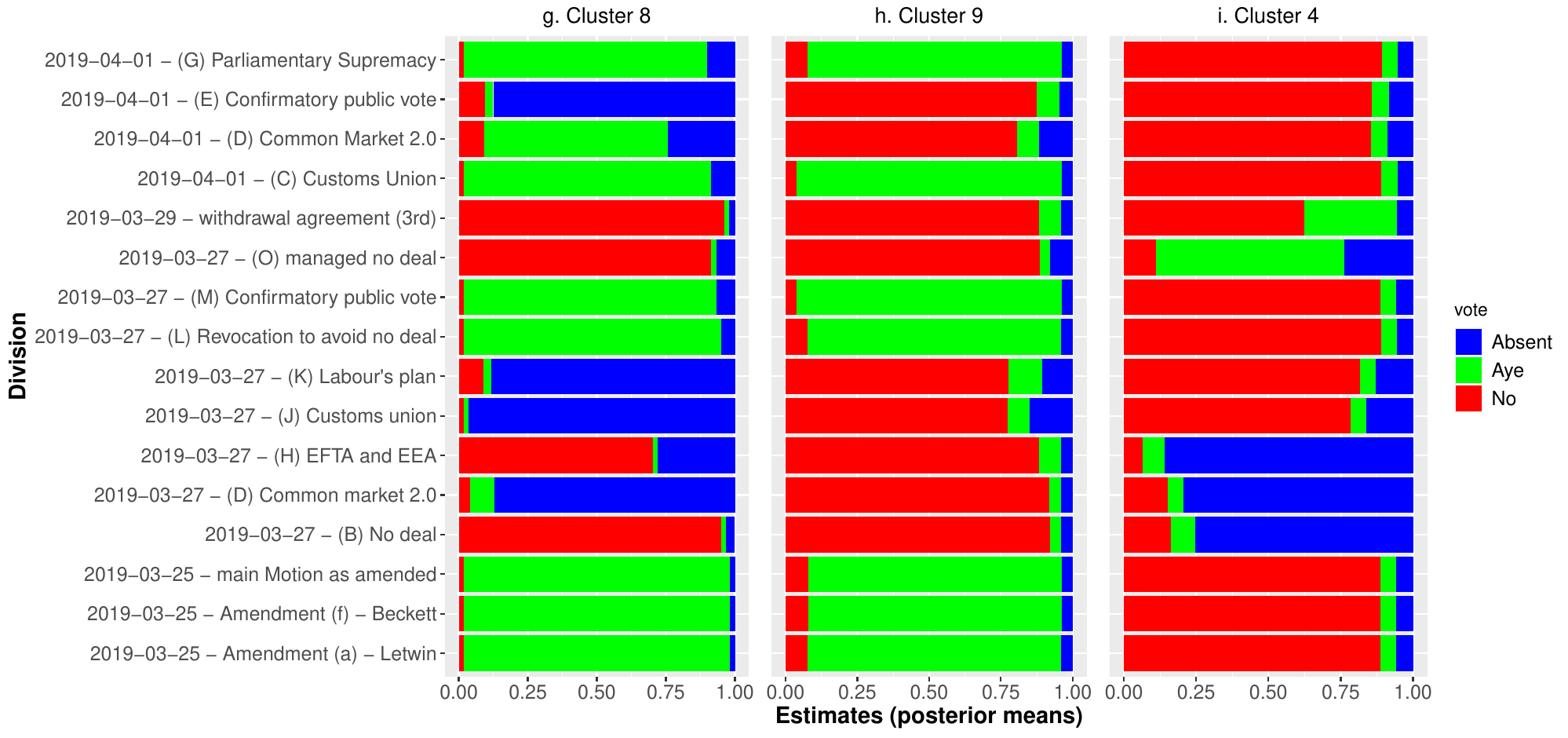}\\
\caption{Vote estimates (posterior means). } 
\label{fig:pm}       

\end{figure}

For interpretation reasons, it is worth mentioning that the log-odds $\eta_g(\mathbf{x})$, $g=1,\dots,10$, are expressed using Cluster 11 as the reference. 
This cluster is the most numerous, with 217 MPs, and is characterized by an extreme pro-Leave (even pro-no-deal) position, as shown in Figure~\ref{fig:pm}.a. Looking at the composition, 211 out of these 217 MPs are conservatives, including Mr John Baron, proposer of the no-deal, Mr Marcus Fysh, proposer of the managed no-deal, and Mr Boris Johnson, who subsequently became prime minister, on July 24th, 2019.
Cluster 2 is (mostly) conservative as well, but the opinion of its MPs looks more heterogenous with respect to Clusters 11. In particular, a clear anti no-deal position emerges in Figure \ref{fig:pm}.b. This might be due to the characteristics of the MPs themselves and the constituencies they represent. In fact, although Conservatives, most of these MPs come from constituencies that expressed a pro-Remain position. 
Both Mr Nicholas Boles and Mr Kenneth Clarke, proposers of the two couples of divisions for a customs union with the EU (``Common market 2.0" and, indeed, ``Customs union"), belong to this group. 
Cluster 7 includes all of the MPs of the Cabinet (of the time), including Mrs Theresa May. A high rate of abstensionism is present in this group, apart from the Letwin-Beckett amendment and the third meaningful vote; see Figure \ref{fig:pm}.c. Covariates seem to have no effect here. 

\begin{figure}

\includegraphics[width=\textwidth]{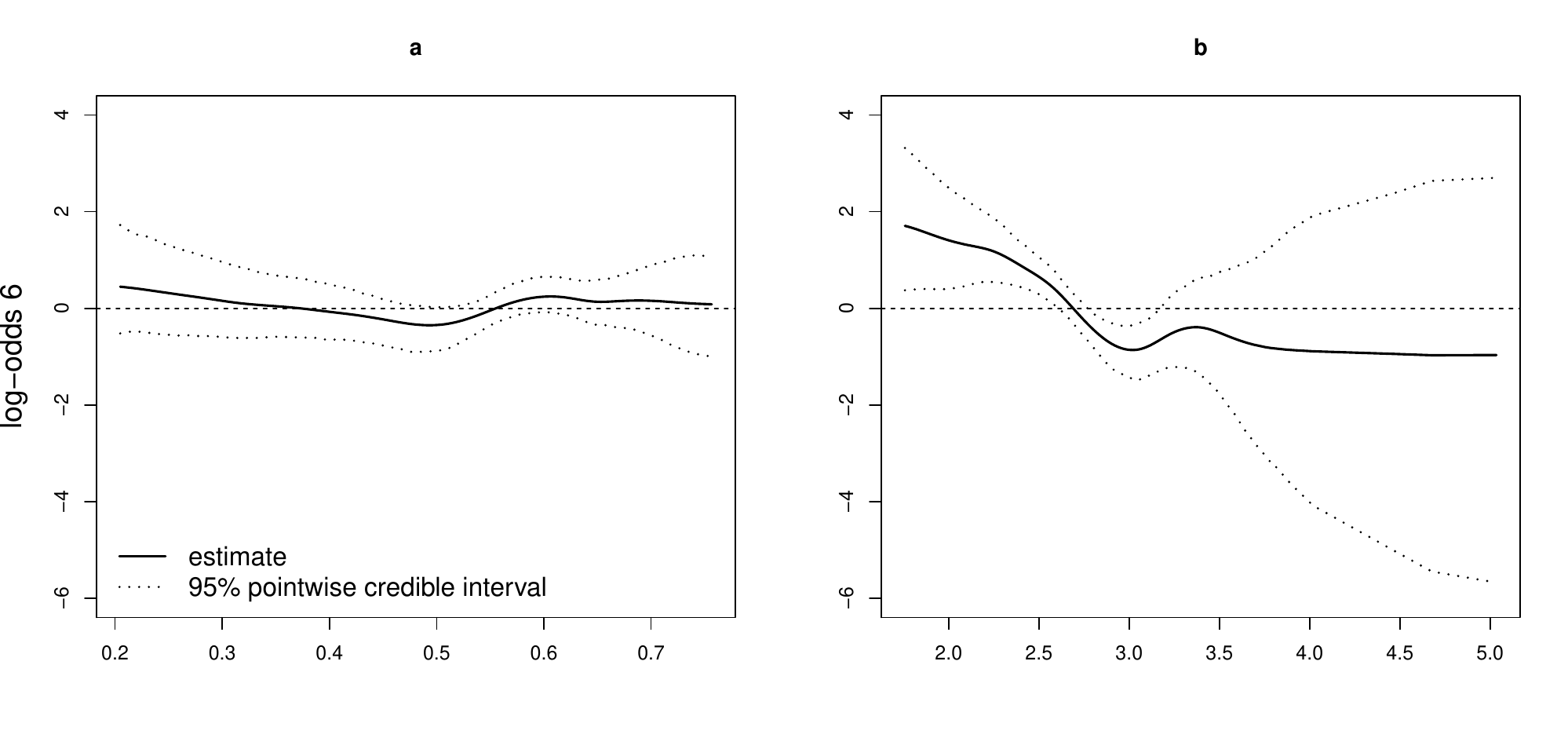}\\
\includegraphics[width=\textwidth]{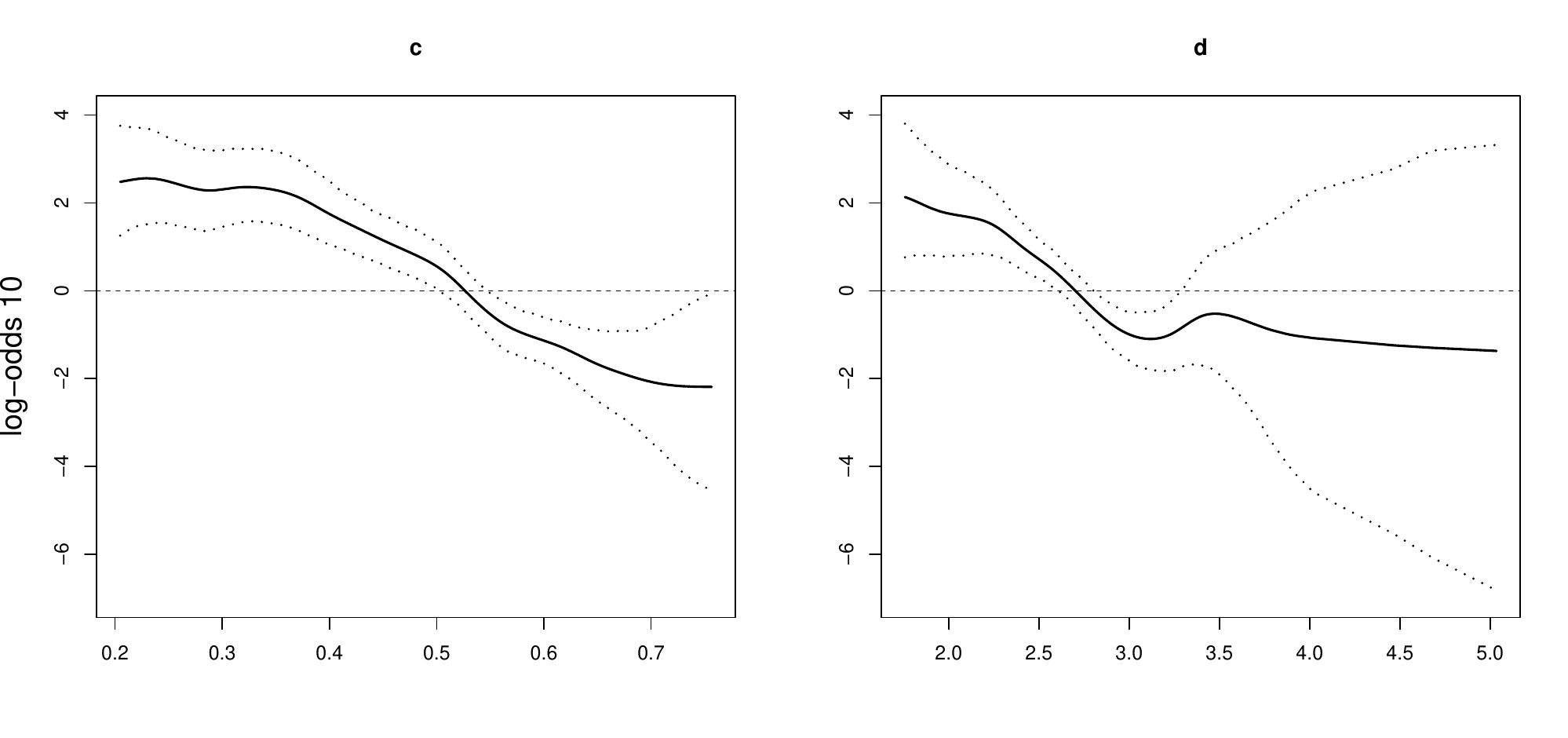}\\
\includegraphics[width=\textwidth]{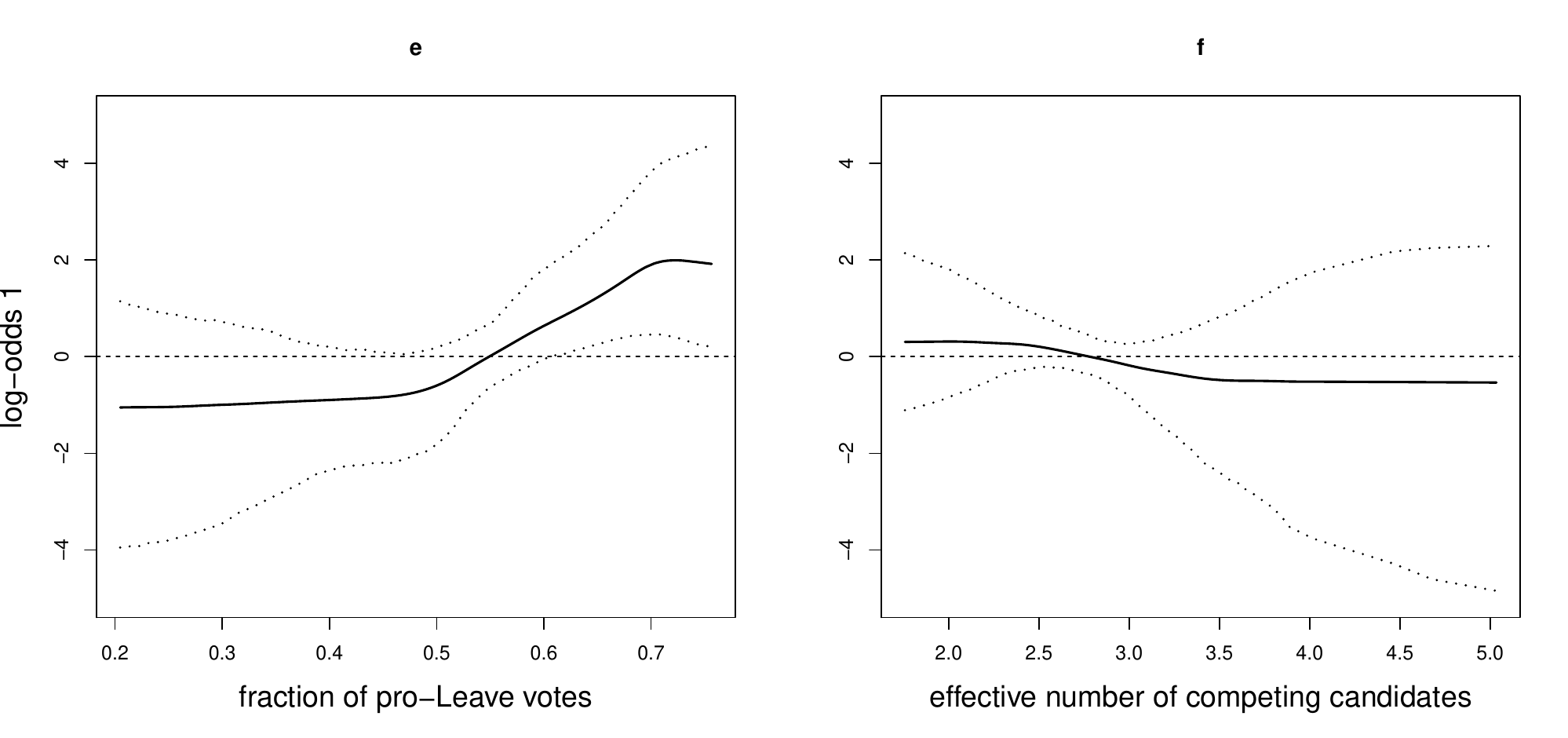}\\
\caption{Estimated smooth effects (with pointwise credible intervals) of the covariates on the log-odds of the mixture weights for Clusters 6 (top panels), 10 (middle panels), 1 (bottom panels); reference: Cluster 11.} 
\label{fig:sf.lab}       

\end{figure}

On the other end, Cluster 6 and 10 are characterized  by a pro-Remain position towards Brexit (see Figure~\ref{fig:pm}.d and Figure~\ref{fig:pm}.e). 
Both are mostly made up by Labour and represent constituencies where the competition is usually concentrated around the two leading parties. 
In fact, the probability for an MP to belong to this cluster is higher if the effective number of competing candidates is lower than 3: this threshold (nonlinear) effect is represented in 
Figure~\ref{fig:sf.lab}.b and Figure~\ref{fig:sf.lab}.d.
The main difference between these two clusters is due to the opinion towards Brexit of the respectively represented constituencies. 
In particular, as it can be seen in Figure~\ref{fig:sf.lab}.a and Figure~\ref{fig:sf.lab}.c, the fraction of leave votes has a mild nonlinear effect on the probability of belonging to Cluster 10, while it has no effect for Cluster 6. 
This seems to be reflected by a more extreme position of the MPs of Cluster 10, especially towards the revocation to avoid no deal (Motions L, G).
The aforementioned Dame Margaret Beckett, first proposer of the confirmatory public vote, as well as of the homonymous amendment, belongs to this group, while the second co-proposers of the confirmatory public vote, Mr Peter Kyle and Mr Phil Wilson are divided between Clusters 10 and 6, respectively.
Cluster 6 is also characterized by the presence of Mr Jeremy Corbyn, proposer of the Labour's alternative plan for Brexit, and leader of the Labour Party until the 2019 United Kingdom general election.
There is a third Labour group, Cluster 1, which represents the pro-Leave minority of the party. 
More precisely, Figure \ref{fig:sf.lab}.e and  shows that the MPs belonging to this cluster distinguish themselves from party members assigned to different clusters because they were elected in the few pro-Leave Labour constituencies. 
This seems to affect their opinion in terms of divisions, which sensibly departs from the party line. 
However, this cluster does not look homogeneous, but rather divided into two fractions. 
In particular Figure \ref{fig:pm}.f shows five divisions where the ``no" fraction is close to 50\%. 

\begin{figure}

\includegraphics[width=\textwidth]{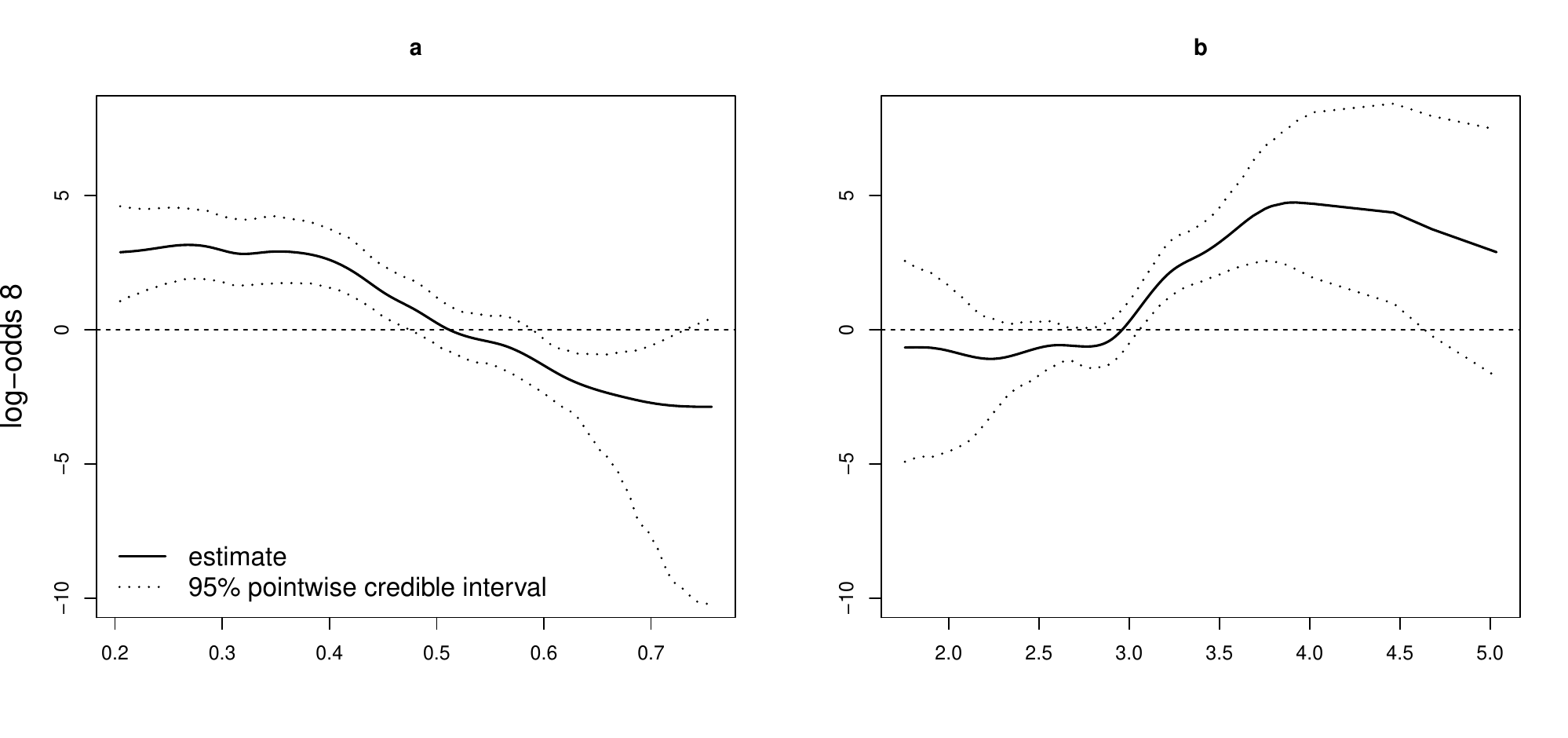}\\
\includegraphics[width=\textwidth]{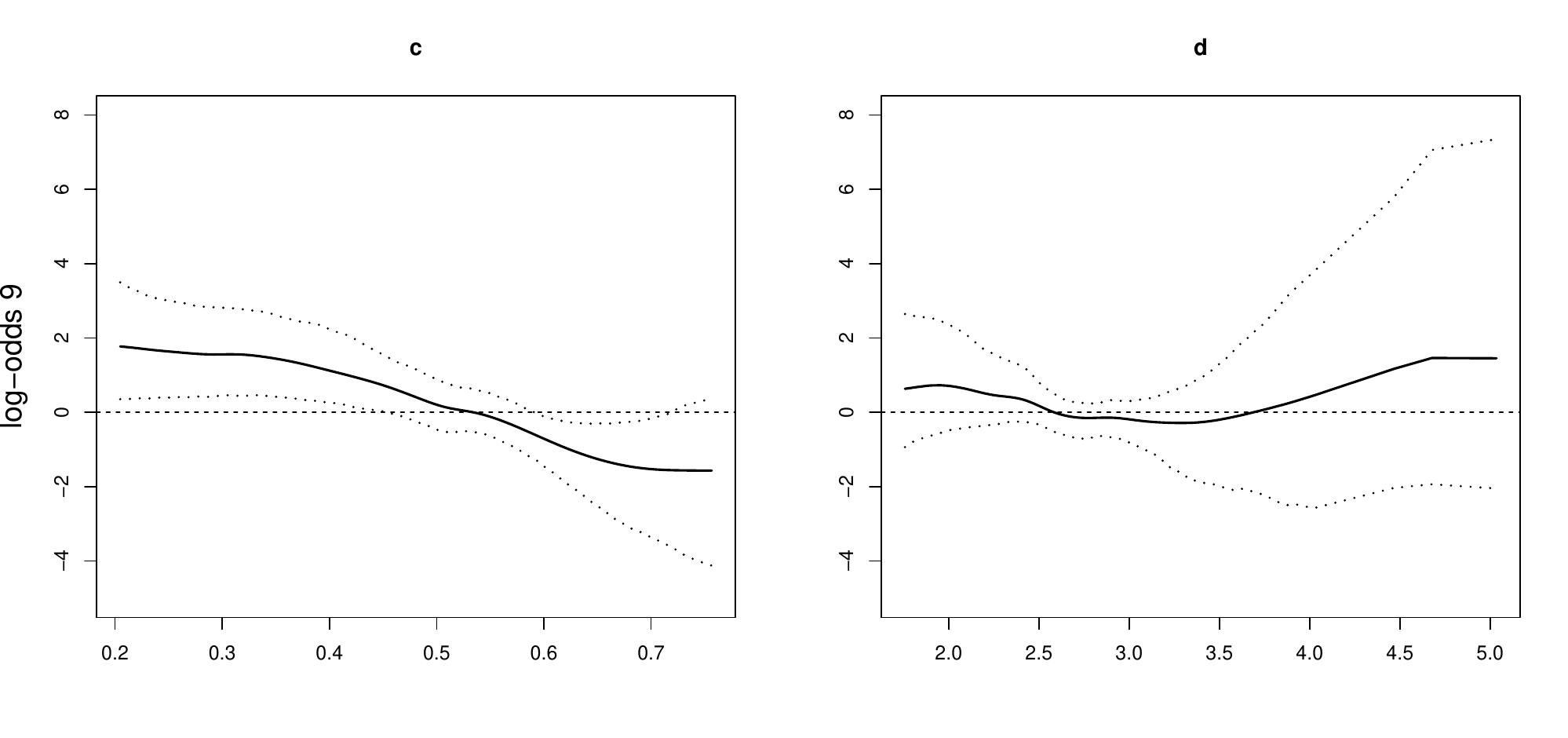}\\
\includegraphics[width=\textwidth]{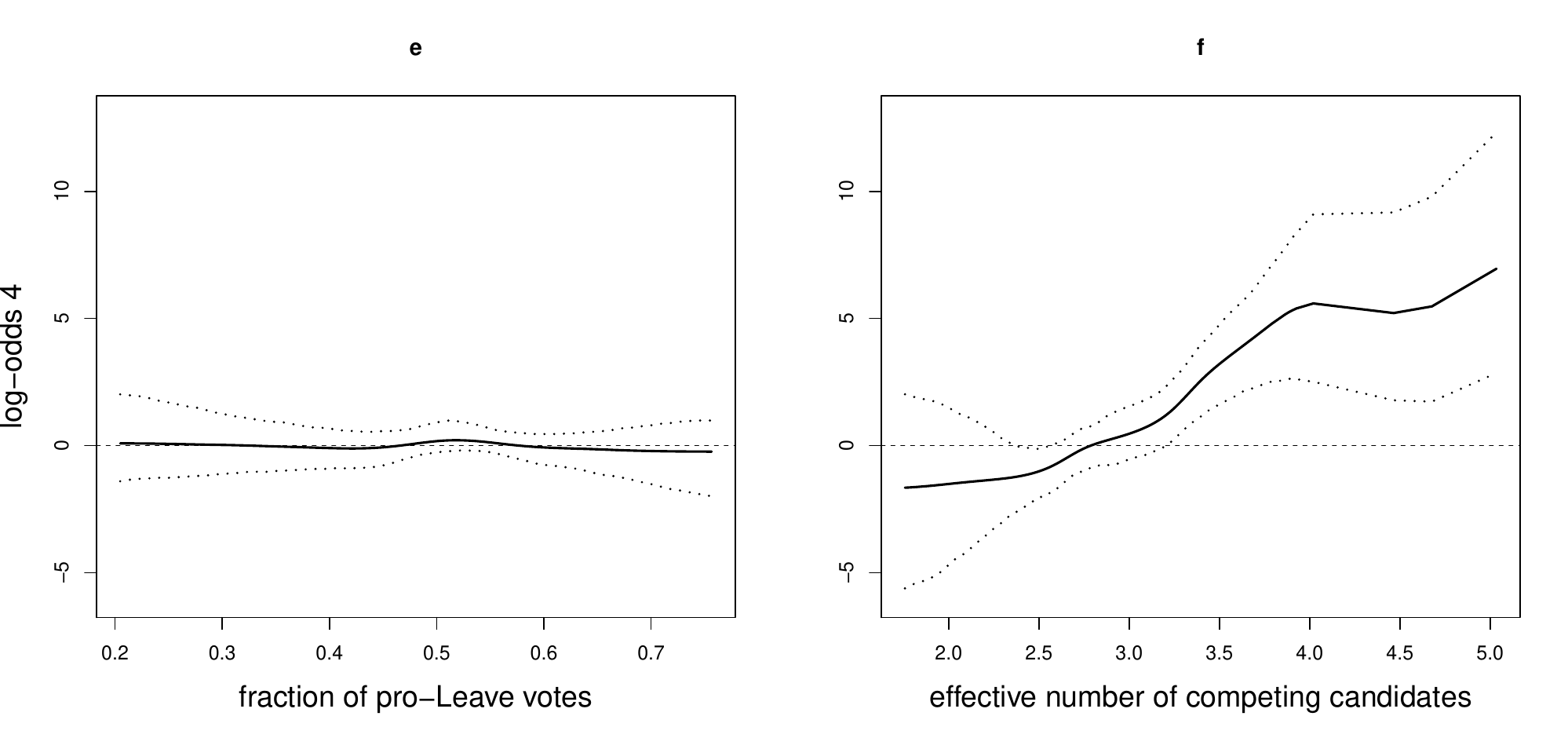}\\
\caption{Estimated smooth effects (with pointwise credible intervals) of the covariates on the log-odds of the mixture weights for Clusters 8 (top panel), 9 (middle panel), 4 (bottom panel); reference: Cluster 11.} 
\label{fig:sf.minor}       

\end{figure}

Cluster 8 is the most heterogeneous in terms of political party membership. 
In fact, 47 MPs out of 54 belonging to the Scottish National Party (including Ms Joanna Cherry, who proposed the revocation of Article 50 to avoid no deal), Plaid Cymru and the Liberal Democrats are in this group, together with seven more MPs belonging to three other different parties. 
Since there is no area where the amount of votes for both the two leading parties (Conservatives and Labour) is negligible, the constituencies represented in this group are usually characterized by the presence of at least 3 political parties, i.e. the one that actually won plus the two leading parties. 
Hence, in this case the effective number of competing candidates has the opposite nonlinear effect on the probability of belonging to these groups (see Figure~\ref{fig:sf.minor}.b), if compared to the ones observed for Clusters 6 and 10. 
A mild nonlinearity appears also in Figure \ref{fig:sf.minor}.a (similar to the one observed for Cluster 10), representing the effect of the opinion about Brexit in the constituency. 
In particular, the probability for an MP to be assigned to Cluster 8 decreases as the fraction of Leave votes in his constituency increases. 
In fact, most of the constituencies represented in this group are known to be pro-Remain, and so are the aforementioned parties, as confirmed by the posterior means of votes shown in Figure~\ref{fig:pm}.g. 
Cluster 9 includes a large portion of the Independents group, made up by MPs who quitted the party they were elected with, and who subsequently founded Change UK: a centrist, pro-European Union political party officially registered on April 15, 2019, and dissolved eight months later, after the 2019 general election. 
Their voting behaviour seems to reflect the pro-Remain position of the constituencies where they were elected from (Figures \ref{fig:sf.minor}.c and \ref{fig:pm}.h).
Finally, Cluster 4 is the last remaining group which is not mostly composed by MPs belonging to the leading parties, being mostly made up by the whole Democratic Unionist Party. A threshold effect of the number of competing candidates similar (slightly smoother) to the one observed for Cluster 8 is present (Figure~\ref{fig:sf.minor}.f). Members of Cluster 4 tend to abstein or vote against all the motions but the one for the managed no deal; see Figure~\ref{fig:pm}.i.

\section{Conclusions}
\label{sec:7}
The analysis of the determinants of cluster membership of the United Kingdom MPs reported in this article shows that political party membership is not sufficient to explain their position towards Brexit, coherently with the results obtained by \citet{intal2021dissent} via network analysis.
More specifically, both the safeness of seat (in terms of effective number of competing candidates at the previous general election) and the share of leave vote in the constituency have a strong influence. This result is in line with the conclusions drawn by \citet{aidt2021meaningful}, who observe that the effect of constituency preferences is significantly stronger in marginal constituencies.
Thus, the inclusion of these covariates in the study helps detecting some homogeneous sub-groups of MPs within the two major parties.
Furthermore, the proposed methodology allows to catch nonlinearity in the effect of this concomitant information on the log-odds of cluster membership of the MPs, through the specification of  a new class of semiparametric latent class models. This approach exploits an additive structure for the mixing proportions, and resorts to spline functions for approximating the smooth effect of the covariates. 
Parameter estimation is based on a formal Bayesian approach through MCMC machinery.
A similar result, in principle, could be emulated also through a fully parametric approach, e.g. by considering a monomial set of bases to represent the map between component probabilities and covariates. However, resorting to a parametric representation, flexible enough to catch nonlinearity, would require some arbitrary choices, such as the maximum degree for monomial bases, or the definition of an automatic selection criterion, wheras the approach suggested in this paper bypasses this issue by controlling flexibility through the variance parameters of the spline coefficients, following \citet{lang2004bayesian}. 

Although the results shown in this paper seem encouraging, there are some issues that might deserve further investigation. First of all, divisions are assumed to be independent (conditional on cluster membership), but different specifications might be possible for the multivariate discrete case, e.g. based on the so-called underlying random variables (URV) approaches \citep{ranalli2017mixture}, which, however, require the manifest variables to be ordinal. Moreover, even though the variables object of this study are categorical, the proposed methodology can be adapted to any other type of response variables (e.g. continuous), by choosing an appropriate form for the component density $f(\mathbf{y}|\boldsymbol{\theta}_g)$. Indeed, it is worth noting that the main methodological result presented in this paper regards the possibility to provide a flexible specification of the component weights, whose implementation is not influenced by the form of the component distribution in practice. 
One of the main advantages of the proposed MCMC algorithm is the absence of Metropolis-Hastings steps. On the other hand, the use of mixture of Gaussians to approximate the logistic distribution introduces an additional latent variable that can increase the computational burden. Furthermore, the implemented Gibbs sampler requires the number of components as input. If this quantity is unknown, one needs to estimate it by running the algorithm many times with different inputs, which might be time consuming, especially when the ``true" value is high. One solution could be incorporating the choice of the number of components within the algorithm itself. As observed in the simulation studies (reported in the supplementary materials), it might happen that the proposed MCMC algorithm converges to a solution that is characterised by empty components. This peculiar behaviour could be exploited to devise a strategy similar to the one proposed by \citet{malsiner2016model}, which circumvents the issue of choosing the optimal value for $G$ and focuses the attention on the posterior distribution of the number of non-empty components, by combining a large value for $G$ with appropriate prior distributions. Alternatively, a reversible jump MCMC algorithm could be exploited \citep{richardson1997bayesian}, by designing appropriate dimension-changing moves (such as split-and-merge moves and birth-and-death moves).
 

\newpage

\bibliographystyle{Chicago}

\bibliography{Cladag2019bib}

\begin{thebibliography}{}

\bibitem[\protect\citeauthoryear{Aidt, Grey, and Savu}{Aidt
  et~al.}{2021}]{aidt2021meaningful}
Aidt, T., F.~Grey, and A.~Savu (2021).
\newblock {The Meaningful Votes: Voting on Brexit in the British House of
  Commons}.
\newblock {\em Public Choice\/}~{\em 186\/}(3), 587--617.

\bibitem[\protect\citeauthoryear{Apostolova, Audickas, Baker, Bate, Cracknell,
  Dempsey, Hawkins, McInnes, Rutherford, and Uberoi}{Apostolova
  et~al.}{2017}]{apostolova2017general}
Apostolova, V., L.~Audickas, C.~Baker, A.~Bate, R.~Cracknell, N.~Dempsey,
  O.~Hawkins, R.~McInnes, T.~Rutherford, and E.~Uberoi (2017).
\newblock General election 2017: results and analysis.
\newblock Briefing Paper no CBP 7979.

\bibitem[\protect\citeauthoryear{B{\"o}hning}{B{\"o}hning}{1999}]{bohning1999computer}
B{\"o}hning, D. (1999).
\newblock {\em Computer-assisted analysis of mixtures and applications:
  meta-analysis, disease mapping and others}, Volume~81.
\newblock CRC Press.

\bibitem[\protect\citeauthoryear{Bouveyron, Celeux, Murphy, and
  Raftery}{Bouveyron et~al.}{2019}]{bouveyron2019model}
Bouveyron, C., G.~Celeux, T.~B. Murphy, and A.~E. Raftery (2019).
\newblock {\em Model-based clustering and classification for data science: with
  applications in R}, Volume~50.
\newblock Cambridge University Press.

\bibitem[\protect\citeauthoryear{Brezger and Lang}{Brezger and
  Lang}{2006}]{brezger2006generalized}
Brezger, A. and S.~Lang (2006).
\newblock Generalized structured additive regression based on {B}ayesian
  {P}-splines.
\newblock {\em Computational Statistics \& Data Analysis\/}~{\em 50\/}(4),
  967--991.

\bibitem[\protect\citeauthoryear{Chamroukhi}{Chamroukhi}{2015}]{chamroukhi2015non}
Chamroukhi, F. (2015).
\newblock Non-normal mixtures of experts.
\newblock arXiv preprint arXiv:1506.06707.

\bibitem[\protect\citeauthoryear{Dayton and Macready}{Dayton and
  Macready}{1988}]{dayton1988concomitant}
Dayton, C.~M. and G.~B. Macready (1988).
\newblock Concomitant-variable latent-class models.
\newblock {\em Journal of the American Statistical Association\/}~{\em
  83\/}(401), 173--178.

\bibitem[\protect\citeauthoryear{DeSarbo and Cron}{DeSarbo and
  Cron}{1988}]{desarbo1988maximum}
DeSarbo, W.~S. and W.~L. Cron (1988).
\newblock A maximum likelihood methodology for clusterwise linear regression.
\newblock {\em Journal of Classification\/}~{\em 5\/}(2), 249--282.

\bibitem[\protect\citeauthoryear{Erosheva, Fienberg, and Joutard}{Erosheva
  et~al.}{2007}]{erosheva2007describing}
Erosheva, E.~A., S.~E. Fienberg, and C.~Joutard (2007).
\newblock Describing disability through individual-level mixture models for
  multivariate binary data.
\newblock {\em The Annals of Applied Statistics\/}~{\em 1\/}(2), 502--537.

\bibitem[\protect\citeauthoryear{Everitt and Hand}{Everitt and
  Hand}{1981}]{everitt1981finite}
Everitt, B.~S. and D.~J. Hand (1981).
\newblock {\em Finite mixture distributions}.
\newblock CRC Press.

\bibitem[\protect\citeauthoryear{Fr{\"u}hwirth-Schnatter}{Fr{\"u}hwirth-Schnatter}{2006}]{fruhwirth2006finite}
Fr{\"u}hwirth-Schnatter, S. (2006).
\newblock {\em Finite mixture and {M}arkov switching models}.
\newblock Springer.

\bibitem[\protect\citeauthoryear{Fr{\"u}hwirth-Schnatter and
  Fr{\"u}hwirth}{Fr{\"u}hwirth-Schnatter and
  Fr{\"u}hwirth}{2010}]{fruhwirth2010data}
Fr{\"u}hwirth-Schnatter, S. and R.~Fr{\"u}hwirth (2010).
\newblock Data augmentation and {MCMC} for binary and multinomial logit models.
\newblock In {\em Statistical modelling and regression structures}, pp.\
  111--132. Springer.

\bibitem[\protect\citeauthoryear{Fr{\"u}hwirth-Schnatter, Pamminger, Weber, and
  Winter-Ebmer}{Fr{\"u}hwirth-Schnatter et~al.}{2012}]{fruhwirth2012labor}
Fr{\"u}hwirth-Schnatter, S., C.~Pamminger, A.~Weber, and R.~Winter-Ebmer
  (2012).
\newblock Labor market entry and earnings dynamics: {B}ayesian inference using
  mixtures-of-experts {M}arkov chain clustering.
\newblock {\em Journal of Applied Econometrics\/}~{\em 27\/}(7), 1116--1137.

\bibitem[\protect\citeauthoryear{Geman and Geman}{Geman and
  Geman}{1984}]{geman1984stochastic}
Geman, S. and D.~Geman (1984).
\newblock Stochastic relaxation, {G}ibbs distributions, and the {B}ayesian
  restoration of images.
\newblock {\em IEEE Transactions on Pattern Analysis and Machine
  Intelligence\/}~{\em 6\/}(6), 721--741.

\bibitem[\protect\citeauthoryear{Gormley and Fr{\"u}hwirth-Schnatter}{Gormley
  and Fr{\"u}hwirth-Schnatter}{2019}]{gormley2019mixture}
Gormley, I.~C. and S.~Fr{\"u}hwirth-Schnatter (2019).
\newblock Mixture of experts models.
\newblock In {\em Handbook of Mixture Analysis}, pp.\  271--307. CRC Press.

\bibitem[\protect\citeauthoryear{Gormley and Murphy}{Gormley and
  Murphy}{2008}]{gormley2008mixture}
Gormley, I.~C. and T.~B. Murphy (2008).
\newblock A mixture of experts model for rank data with applications in
  election studies.
\newblock {\em The Annals of Applied Statistics\/}~{\em 2\/}(4), 1452--1477.

\bibitem[\protect\citeauthoryear{Gormley and Murphy}{Gormley and
  Murphy}{2010}]{gormley2010mixture}
Gormley, I.~C. and T.~B. Murphy (2010).
\newblock A mixture of experts latent position cluster model for social network
  data.
\newblock {\em Statistical Methodology\/}~{\em 7\/}(3), 385--405.

\bibitem[\protect\citeauthoryear{Gormley and Murphy}{Gormley and
  Murphy}{2011}]{gormley2011mixture}
Gormley, I.~C. and T.~B. Murphy (2011).
\newblock Mixture of experts modelling with social science applications.
\newblock In {\em Mixtures: Estimation and Applications}, pp.\  101--121. Wiley
  Online Library.

\bibitem[\protect\citeauthoryear{Green and Silverman}{Green and
  Silverman}{1993}]{green1993nonparametric}
Green, P.~J. and B.~W. Silverman (1993).
\newblock {\em Nonparametric regression and generalized linear models: a
  roughness penalty approach}.
\newblock Chapman and Hall/CRC.

\bibitem[\protect\citeauthoryear{Hanretty}{Hanretty}{2017}]{hanretty2017areal}
Hanretty, C. (2017).
\newblock Areal interpolation and the {UK}'s referendum on {EU} membership.
\newblock {\em Journal of Elections, Public Opinion and Parties\/}~{\em
  27\/}(4), 466--483.

\bibitem[\protect\citeauthoryear{Hastie, Tibshirani, and Friedman}{Hastie
  et~al.}{2009}]{hastie2009elements}
Hastie, T., R.~Tibshirani, and J.~Friedman (2009).
\newblock {\em The elements of statistical learning: data mining, inference,
  and prediction}.
\newblock Springer.

\bibitem[\protect\citeauthoryear{Hastie and Tibshirani}{Hastie and
  Tibshirani}{1990}]{hastie1990generalized}
Hastie, T.~J. and R.~J. Tibshirani (1990).
\newblock {\em Generalized additive models}, Volume~43.
\newblock CRC press.

\bibitem[\protect\citeauthoryear{Intal and Yasseri}{Intal and
  Yasseri}{2021}]{intal2021dissent}
Intal, C. and T.~Yasseri (2021).
\newblock {Dissent and rebellion in the House of Commons: A social network
  analysis of Brexit-related divisions in the 57th Parliament}.
\newblock {\em Applied Network Science\/}~{\em 6\/}(1), 1--12.

\bibitem[\protect\citeauthoryear{Jacobs, Jordan, Nowlan, and Hinton}{Jacobs
  et~al.}{1991}]{jacobs1991adaptive}
Jacobs, R.~A., M.~I. Jordan, S.~J. Nowlan, and G.~E. Hinton (1991).
\newblock Adaptive mixtures of local experts.
\newblock {\em Neural computation\/}~{\em 3\/}(1), 79--87.

\bibitem[\protect\citeauthoryear{Jordan and Jacobs}{Jordan and
  Jacobs}{1994}]{jordan1994hierarchical}
Jordan, M.~I. and R.~A. Jacobs (1994).
\newblock Hierarchical mixtures of experts and the {EM} algorithm.
\newblock {\em Neural computation\/}~{\em 6\/}(2), 181--214.

\bibitem[\protect\citeauthoryear{Lang and Brezger}{Lang and
  Brezger}{2004}]{lang2004bayesian}
Lang, S. and A.~Brezger (2004).
\newblock Bayesian {P}-splines.
\newblock {\em Journal of Computational and Graphical Statistics\/}~{\em
  13\/}(1), 183--212.

\bibitem[\protect\citeauthoryear{Lindsay}{Lindsay}{1995}]{lindsay1995mixture}
Lindsay, B.~G. (1995).
\newblock {\em Mixture models: theory, geometry and applications}.
\newblock IMS.

\bibitem[\protect\citeauthoryear{MacArthur}{MacArthur}{1965}]{macarthur1965patterns}
MacArthur, R.~H. (1965).
\newblock Patterns of species diversity.
\newblock {\em Biological reviews\/}~{\em 40\/}(4), 510--533.

\bibitem[\protect\citeauthoryear{Malsiner-Walli, Fr{\"u}hwirth-Schnatter, and
  Gr{\"u}n}{Malsiner-Walli et~al.}{2016}]{malsiner2016model}
Malsiner-Walli, G., S.~Fr{\"u}hwirth-Schnatter, and B.~Gr{\"u}n (2016).
\newblock Model-based clustering based on sparse finite {G}aussian mixtures.
\newblock {\em Statistics and Computing\/}~{\em 26}, 303--324.

\bibitem[\protect\citeauthoryear{McLachlan and Basford}{McLachlan and
  Basford}{1988}]{mclachlan1988mixture}
McLachlan, G.~J. and K.~E. Basford (1988).
\newblock {\em Mixture models: Inference and applications to clustering}.
\newblock New York: Dekker.

\bibitem[\protect\citeauthoryear{McLachlan and Peel}{McLachlan and
  Peel}{2004}]{mclachlan2004finite}
McLachlan, G.~J. and D.~Peel (2004).
\newblock {\em Finite mixture models}.
\newblock John Wiley \& Sons.

\bibitem[\protect\citeauthoryear{McNicholas}{McNicholas}{2016}]{mcnicholas2016mixture}
McNicholas, P.~D. (2016).
\newblock {\em Mixture model-based classification}.
\newblock CRC press.

\bibitem[\protect\citeauthoryear{Mengersen, Robert, and Titterington}{Mengersen
  et~al.}{2011}]{mengersen2011mixtures}
Mengersen, K.~L., C.~Robert, and M.~Titterington (2011).
\newblock {\em Mixtures: Estimation and Applications}.
\newblock John Wiley \& Sons.

\bibitem[\protect\citeauthoryear{Metropolis, Rosenbluth, Rosenbluth, Teller,
  and Teller}{Metropolis et~al.}{1953}]{metropolis1953equation}
Metropolis, N., A.~W. Rosenbluth, M.~N. Rosenbluth, A.~H. Teller, and E.~Teller
  (1953).
\newblock Equation of state calculations by fast computing machines.
\newblock {\em The Journal of Chemical Physics\/}~{\em 21\/}(6), 1087--1092.

\bibitem[\protect\citeauthoryear{Mollica and Tardella}{Mollica and
  Tardella}{2017}]{mollica2017bayesian}
Mollica, C. and L.~Tardella (2017).
\newblock Bayesian {P}lackett--{L}uce mixture models for partially ranked data.
\newblock {\em Psychometrika\/}~{\em 82\/}(2), 442--458.

\bibitem[\protect\citeauthoryear{Murphy and Murphy}{Murphy and
  Murphy}{2020}]{murphy2019gaussian}
Murphy, K. and T.~B. Murphy (2020).
\newblock Gaussian parsimonious clustering models with covariates and a noise
  component.
\newblock {\em Advances in Data Analysis and Classification\/}~{\em 14},
  293--325.

\bibitem[\protect\citeauthoryear{Nguyen and Chamroukhi}{Nguyen and
  Chamroukhi}{2018}]{nguyen2018practical}
Nguyen, H.~D. and F.~Chamroukhi (2018).
\newblock Practical and theoretical aspects of mixture-of-experts modeling: An
  overview.
\newblock {\em Wiley Interdisciplinary Reviews: Data Mining and Knowledge
  Discovery\/}~{\em 8\/}(4), e1246.

\bibitem[\protect\citeauthoryear{Nguyen and McLachlan}{Nguyen and
  McLachlan}{2016}]{nguyen2016laplace}
Nguyen, H.~D. and G.~J. McLachlan (2016).
\newblock Laplace mixture of linear experts.
\newblock {\em Computational Statistics \& Data Analysis\/}~{\em 93}, 177--191.

\bibitem[\protect\citeauthoryear{Odell}{Odell}{2017}]{hansard}
Odell, E. (2017).
\newblock {\em {hansard}: Provides Easy Downloading Capabilities for the UK
  Parliament API}.
\newblock R package version 0.8.0.

\bibitem[\protect\citeauthoryear{Quandt}{Quandt}{1972}]{quandt1972new}
Quandt, R.~E. (1972).
\newblock A new approach to estimating switching regressions.
\newblock {\em Journal of the American Statistical Association\/}~{\em
  67\/}(338), 306--310.

\bibitem[\protect\citeauthoryear{Raftery, Newton, Satagopan, and
  Krivitsky}{Raftery et~al.}{2007}]{RafEtAl2007}
Raftery, A., M.~Newton, J.~Satagopan, and P.~Krivitsky (2007).
\newblock {E}stimating the integrated likelihood via posterior simulation using
  the harmonic mean identity.
\newblock In {\em Bayesian Statistics 8}, pp.\  371--416. Oxford University
  Press.

\bibitem[\protect\citeauthoryear{Ranalli and Rocci}{Ranalli and
  Rocci}{2017}]{ranalli2017mixture}
Ranalli, M. and R.~Rocci (2017).
\newblock Mixture models for mixed-type data through a composite likelihood
  approach.
\newblock {\em Computational Statistics \& Data Analysis\/}~{\em 110}, 87--102.

\bibitem[\protect\citeauthoryear{Richardson and Green}{Richardson and
  Green}{1997}]{richardson1997bayesian}
Richardson, S. and P.~J. Green (1997).
\newblock On {B}ayesian analysis of mixtures with an unknown number of
  components (with discussion).
\newblock {\em Journal of the Royal Statistical Society: Series B (statistical
  methodology)\/}~{\em 59\/}(4), 731--792.

\bibitem[\protect\citeauthoryear{Rue and Held}{Rue and
  Held}{2005}]{rue2005gaussian}
Rue, H. and L.~Held (2005).
\newblock {\em Gaussian Markov random fields: theory and applications}.
\newblock Chapman and Hall/CRC.

\bibitem[\protect\citeauthoryear{Shannon}{Shannon}{1948}]{shannon1948entropy}
Shannon, C.~E. (1948).
\newblock A mathematical theory of communication.
\newblock {\em The Bell System Technical Journal\/}~{\em 27\/}(3), 379--423.

\bibitem[\protect\citeauthoryear{Spirling and Quinn}{Spirling and
  Quinn}{2010}]{spirling2010identifying}
Spirling, A. and K.~Quinn (2010).
\newblock {Identifying intraparty voting blocs in the UK House of Commons}.
\newblock {\em Journal of the American Statistical Association\/}~{\em
  105\/}(490), 447--457.

\bibitem[\protect\citeauthoryear{Tang and Qu}{Tang and
  Qu}{2016}]{tang2016mixture}
Tang, X. and A.~Qu (2016).
\newblock Mixture modeling for longitudinal data.
\newblock {\em Journal of Computational and Graphical Statistics\/}~{\em
  25\/}(4), 1117--1137.

\bibitem[\protect\citeauthoryear{Titterington, Smith, and Makov}{Titterington
  et~al.}{1985}]{titterington1985statistical}
Titterington, D.~M., A.~F. Smith, and U.~E. Makov (1985).
\newblock {\em Statistical analysis of finite mixture distributions}.
\newblock Wiley,.

\bibitem[\protect\citeauthoryear{Wang, Puterman, Cockburn, and Le}{Wang
  et~al.}{1996}]{wang1996mixed}
Wang, P., M.~L. Puterman, I.~Cockburn, and N.~Le (1996).
\newblock Mixed {P}oisson regression models with covariate dependent rates.
\newblock {\em Biometrics\/}~{\em 52\/}(2), 381--400.

\bibitem[\protect\citeauthoryear{Wood}{Wood}{2017}]{wood2017generalized}
Wood, S.~N. (2017).
\newblock {\em Generalized additive models: an introduction with R}.
\newblock CRC press.

\end{thebibliography}

\end{document}